\documentclass[twocolumn,aps,pra,superscriptaddress,showpacs,floatfix]{revtex4}
\usepackage{graphicx}
\usepackage{amsmath}
\usepackage{amssymb}
\usepackage{amsthm}

\newtheorem*{prop*}{Proposition}

\begin{document}

\title{Quantum Fisher Information Flow in Non-Markovian Processes of Open
Systems}
\author{Xiao-Ming Lu}
\affiliation{Zhejiang Institute of Modern Physics, Department of Physics, Zhejiang
University, Hangzhou 310027, China}
\author{Xiaoguang Wang}
\email{xgwang@zimp.zju.edu.cn}
\affiliation{Zhejiang Institute of Modern Physics, Department of Physics, Zhejiang
University, Hangzhou 310027, China}
\author{C. P. Sun}
\email{suncp@itp.ac.cn}
\affiliation{Institute of Theoretical Physics, Chinese Academy of Sciences, Beijing,
100080, China}

\begin{abstract}
We establish an information-theoretic approach for quantitatively
characterizing the Non-Markovianity of open quantum processes. Here, the
quantum Fisher information (QFI) flow provides a measure to statistically
distinguish Markovian and non-Markovian processes. A basic relation between
the QFI flow and non-Markovianity is unveiled for quantum dynamics of open
systems. For a class of time-local master equations, the exactly-analytic
solution shows that for each fixed time the QFI flow is
decomposed into additive sub-flows according to different dissipative
channels.
\end{abstract}

\maketitle

\section{Introduction} Any system in the realistic world is open since it
inevitably interacts with its environment. The time evolutions of open
systems are simply classified into Markovian and non-Markovian ones
according to the ways to lose energy or information~\cite{BreuerBook}. In
most situations, Markovian process uniquely determines its final steady
state as an thermal equilibrium~\cite{ssb}, which is independent of its
initial one. In this sense a Markovian process is essentially an information
erasure process, thus tends to continuously reduce the distinguishability
between any two initial states~\cite{Breuer2009}.

However, Markovian description for an open quantum system is only an
approximation to most of realistic processes, which are of non-Markovian.
With many recent investigations about non-Markovian dynamics by making use
of various analytical approaches and numerical simulations, a computable
measure of ``Markovianity'' for quantum channels was introduced in Ref.~\cite%
{Wolf2008}. Most recently, it was also recognized that difference between
them can be measured through the continuous increment of the state
distinguishability~\cite{Breuer2009}. Then this increment is intuitively
interpreted as the revival of information flow between the bath and the
system though there no quantitative information measure is utilized.  Based
on this measure of the non-Markovianity, a method for direct measurement of  the
non-Markovian character was proposed~\cite{ZYXu2010}. Another
approach based on entanglement is proposed in Ref.~\cite{Rivas2009}. In this
paper, the quantum Fisher information (QFI) flow are introduced to directly
characterize the non-Markovianity of the quantum dynamics of open systems.

Actually, in the system-plus-bath approach for open quantum systems, the
effective dynamics of the reduced density matrix $\rho$ is induced by
tracing over environment~\cite{BreuerBook}. The simplest reduced dynamics is
the quantum Markovian process described by dynamical semigroups~\cite%
{AlickiBook}. There, the reduced density matrix $\rho$ at time $t+dt$ is
determined completely by the one at time $t$. Contrarily, the general
reduced dynamics may be of non-Markovian when the surrounding environment
may retain a memory of the information about states at earlier times, and
transfer it back to the system to affect its evolution. In this sense the
Markovian process only happens when the environmental correlation time is
relatively short so that memory effects can be neglected. These memory-based
considerations for the Markovian approximation also mean that the
information-theoretical characterizing of the non-Markovianity is a quite
natural fashion. However, it is still an open question that how to treat the
information flow in open quantum systems based on a solid information-theoretic
foundation.

In this paper, we establish such a framework by adopting the QFI flow as
the quantitative measure for the information flow. The
QFI characterizes the statistical distinguishability of reduced density
matrix~\cite{Wootters1981,Braunstein1994}. An intuitive picture of the
memory effect of a non-Markovian behavior then immediately follows from the
dynamic return of the QFI, which is depicted by the inward QFI flow. For a
class of the non-Markovian master equations in time-local forms, we exactly
calculate the information flows. The analytic results show that the total
QFI flow can be decomposed into the split contributions from different
dissipative channels for each fixed time. On the other hand, the QFI plays an essential role in
quantum metrology~\cite{Giovannetti2006}, where the highest precision of
estimating an unknown parameter we may achieve is related to inverse of the
QFI. We point out this QFI flow approach is feasible to work for
understanding the problems of quantum metrology.

\section{Quantum Fisher information in non-Markovian dynamics} We consider
the quantum processes described by the following time-local master equation~%
\cite{Breuer2009,Kossakowski2010}
\begin{equation}
\frac{\partial}{\partial t}\rho(t)=\mathcal{K}(t)\rho(t),  \label{eq:ms}
\end{equation}
where $\mathcal{K}(t)$ is a super-operator acting on the reduced density
matrix $\rho(t)$ as~\cite{LindbladForm,Breuer2004,Kossakowski2010}
\begin{equation}
\mathcal{K}(t)\rho=-i\left[H,\rho\right]+\sum_{i}\gamma_{i}\left[A_{i}\rho
A_{i}^{\dagger}-\frac{1}{2}\left\{ A_{i}^{\dagger}A_{i},\rho\right\} \right],
\label{eq:K_t_Form}
\end{equation}
with $H(t)$ the Hermitian Hamiltonian for the open quantum system
without the couplings to the bath. $\{\cdot,\cdot\}$ denotes the
anti-commutator. If all $\gamma_{i}$ and $A_{i}$ are time independent, and
all $\gamma_{i}$ are positive, equation (\ref{eq:K_t_Form}) is the
conventional master equation of the Lindblad form~\cite{LindbladForm}, which
describes the conventional Markovian process. However, by making use of a
variety of methods, such as the time-convolutionless projection operator
technique~\cite{TCLmethod}, Feynman-Vervon influence functional theory~\cite%
{FVmethod} and some others~\cite{Budini2006}, the parameters $%
\gamma_{i}=\gamma_{i}(t)$ and $A_{i}=A_{i}(t)$ in the time-local master
equation may explicitly depend on time, and $\gamma_{i}$ even may become
negative sometimes. Obviously, the non-Markovian character resides in these
time-dependent coefficients.

Taking some real number $\theta$ in the reduced density matrix $%
\rho(\theta;t)$ as the inference parameter, we write down the QFI~\cite%
{estimationBook}
\begin{equation}
\mathcal{F}(\theta;t):=\mathrm{Tr}\left[L^{2}(\theta;t)\rho(\theta;t)\right],
\label{eq:QFI}
\end{equation}
where $L(\theta;t)$ is the so-called symmetric logarithmic derivative (SLD),
which are Hermitian operators determined by~\cite{estimationBook}
\begin{equation}
\frac{\partial}{\partial\theta}\rho(\theta;t)= \frac{1}{2}\left[%
L(\theta;t)\rho(\theta;t)+\rho(\theta;t)L(\theta;t)\right].  \label{eq:SLD}
\end{equation}
An important essential feature of the QFI is that we can obtain the
achievable lower bound of the mean-square error of unbiased estimators for
the parameter $\theta$ through the quantum Cram\'{e}r-Rao (QCR) theorem
\begin{equation}
\mathrm{Var}(\theta;t)\geq \frac{1}{ M\mathcal{F}(\theta;t)},
\label{eq:QCR}
\end{equation}
where $M$ represents the times of the independent measurements~\cite{estimationBook}.
From the QCR theorem, we can see that the QFI is indeed a measure of a certain
kind of information with respect to the precision of  estimating the inference  parameter.
The relations between the QFI and the statistical distinguishability of $%
\rho(\theta;t)$ and its neighbor has been pointed out in some previous works~%
\cite{estimationBook,Wootters1981,Braunstein1994}.

\emph{Flow of the QFI and its decomposition.---} Here we use the QFI to
characterize the non-Markovianity of the open quantum system by introducing
the QFI flow, which is defined as the change rate $\mathcal{I}:=\partial%
\mathcal{F}/\partial t$ of the QFI. As a central result in this paper, a
proposition about the decomposition of the QFI flow is given as follows:

\begin{prop*}
For an open quantum system described by the time local master  equation~(\ref%
{eq:ms}), the QFI flow  $\mathcal{I}=\sum_{i}\mathcal{I}_{i}$ is explicitly
written as a sum  of subflows $\mathcal{I}_{i}=\gamma_{i}\mathcal{J}_{i}$
with
\begin{equation}
\mathcal{J}_{i}:=-\mathrm{Tr}\left\{ \rho\left[L,A_{i}\right]^{\dagger}\left[%
L,A_{i}\right]\right\} \leq0.  \label{eq:pro}
\end{equation}
\end{prop*}

\emph{Proof.} From the differential of Eq.~(\ref{eq:SLD}) with respect to
time $t$, we have
\begin{equation}
\partial_{t}\partial_{\theta}\rho(\theta)=\frac{1}{2}\left[\dot{L}\rho+L\dot{%
\rho}+\dot{\rho}L+\rho\dot{L}\right].
\end{equation}
It gives
\begin{equation}
\mathrm{Tr}\left[\rho\dot{L}L+\rho L\dot{L}\right]=\mathrm{Tr[}%
2L\partial_{t}\partial_{\theta}\rho(\theta)\mathrm{]-Tr}\left[2\dot{\rho}%
L^{2}\right].
\end{equation}
From the differential of Eq.~(\ref{eq:QFI}) with respect to time $t$, we
obtain the QFI flow as
\begin{equation}
\mathcal{I}=\mathrm{Tr}\left[\mathcal{L}\left(\frac{\partial \rho}{\partial t%
}\right)\right],  \label{eq:dI}
\end{equation}
where the operator $\mathcal{L}:=L(2\partial/\partial\theta-L)$ is defined.
By using the concrete expression of the master equation~(\ref{eq:K_t_Form}),
we split the QFI flow into those individuals corresponding to the different
dissipative channels as $\mathcal{I}=\mathrm{Tr}\left[\mathcal{LK}(t)\rho(t)%
\right]$ or $\mathcal{I}=\sum_{i}\gamma_{i}\mathrm{Tr}[\mathcal{L}(A_{i}\rho
A_{i}^{\dagger})-\frac{1}{2}\mathcal{L}\{ A_{i}^{\dagger}A_{i},\rho\}]$.
After some algebra, we get the decomposition $\mathcal{I}=\sum_{i}\gamma_{i}%
\mathcal{J}_{i}$, where $\mathcal{J}_{i}$ is just given in Eq.~(\ref{eq:pro}%
). It finally proves the proposition.

The above proposition and its proof contain rich implications in physics.
Firstly, the decomposition of the QFI flow corresponding to the different
dissipative channels is due to the linearity of QFI flow equation~(\ref%
{eq:dI}) with respect to $\partial\rho/\partial t$ and the concrete form of
the time-local master equation~(\ref{eq:K_t_Form}). This is not a simple
decomposition since each subflow depends on the whole SLD $L(\theta;t)$,
meanwhile, $L(\theta;t)$ is deduced from $\rho(\theta;t),$ whose evolution
depends on every dissipative channel and the unitary part of the master
equation. So this kind of decomposition does not mean different dissipative
channels are separable to influence the change of the QFI for a period of
time. However, for each fixed time $t>0$, the QFI flow at the present moment
are decomposed into the split contributions from different dissipative
channels. In this sense, we interpret $\mathcal{I}_{i}(t)=\gamma_{i}(t)%
\mathcal{J}_{i}(t)$ as a subflow of the QFI at time $t$ caused by the
dissipative channel described by $A_{i}(t)$ and $\gamma_{i}(t)$. The magnitude
of the QFI subflow is determined by a state-independent factor $\gamma_{i}$
and a state-dependent factor $\mathcal{J}_{i}$.

Secondly, one of the advantages of such decomposition comes from the link
between the direction of each QFI subflow $\mathcal{I}_{i}$ and the sign of
the decay rate $\gamma_{i}$. Because $\mathcal{J}_{i}$ is non-positive, we
conclude that a negative $\gamma_{i}(t)$ implies an inward QFI subflow ($%
\mathcal{I}_{i}>0$), except the trivial case of $\mathcal{J}_{i}=0$. The
temporary appearance of negative decay rates is already considered as the
essential feature of the non-Markovian behaviors~\cite{negativeRate}, here
this is justified through the return of the QFI. For the case that all $%
\gamma_{i}(t)$ are positive, the master equation (\ref{eq:K_t_Form}) describes a so-called
time-dependent Markovian quantum process~\cite%
{Wonderen2000,Maniscalco2004,Wolf2008,Breuer2004}, in which cases, $\mathcal{%
I}$ always decreases. If the total QFI flow $\mathcal{I}(t)$ is positive at
time $t$, it signifies at least one of $\gamma_{i}(t)$ is negative. In such
cases, the QFI flows back to the open system and the non-Markovian behavior
emerges.

Actually, like the trace distance used in Ref.~\cite{Breuer2009}, the
dynamical return of the QFI is linked to the divisibility property of the
dynamical map of quantum processes. If the master equation is of the form (%
\ref{eq:K_t_Form}), the corresponding dynamical map is infinitely divisible
provided that all $\gamma_{i}$ are positive~\cite{Wolf2008-1}. In such
cases, for arbitrary time $t>0$, the dynamical map from time $t$ to $t+dt$
is a completely positive and trace-preserving map. Thus the QFI decreases
during this time interval since the QFI is monotonic with respect to a
completely positive and trace-preserving map~\cite{Fujiwara2001}.

Thirdly, there should be some restriction on the evolution of the QFI. It is
seen from the above proof of the proposition that the coherent part of Eq.~(%
\ref{eq:K_t_Form}), i.e. $-i\left[H(t),\rho(t)\right]$, does not contribute
to the total QFI flow directly. This observation directly leads to the
no-cloning theorem in quantum information, which states that we can not use
unitary operations to evolve the states $|\psi(\theta)\rangle\otimes|0\rangle
$ into $|\psi(\theta)\rangle\otimes|\psi(\theta)\rangle$ as a quantum copy~%
\cite{Wootters1982}. This is because the QFI of the target states is twice
as the one of the source states, due to the additivity of the QFI for the
product states. Besides, if the total system (system plus environment) is
assumed closed and the QFI is only distributed in the system initially, then
the QFI of the reduced density matrix during evolution should be not greater
than the one at the initial time, for the invariance of the QFI under
unitary evolution and the non-increasing of the QFI under partial trace
operation. This restriction should be reflected in the QFI flow obtained
from a proper master equation.
\begin{figure}[!htb]
\begin{centering}
    \includegraphics[scale=0.4]{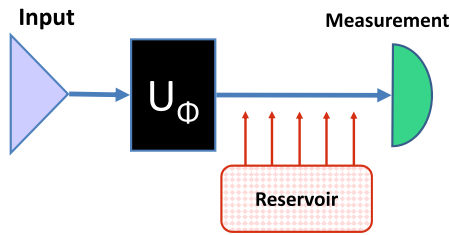}
    \par\end{centering}
\caption{Estimation of parameter $\phi$ in an unitary operation.
After the phase gate operation, the system interacts with a reservoir. The
precision of the estimation is impacted by characteristics of both the
reservoir and the interaction.}
\end{figure}

\section{Two-level system} Now we use an example of two-level system
(qubit) to explicitly illustrate our discovery about the intrinsic relation
between the QFI flow and the non-Markovianity of the open quantum system and
its impaction to parameter estimation. In the quantum metrology context, the
QFI gives a theoretical-achievable limit on the precision when estimating an unknown
parameter, according to the QCR theorem (\ref{eq:QCR}). To estimate the parameter as
precisely as possible, we should optimize input states to maximize the QFI,
and then optimize measurements to achieve the Cram\'{e}r-Rao bound~\cite%
{Giovannetti2006}. However, due to the interaction with environment, the QFI
will change and affect the precision of the parameter estimation.

Here, the QFI-based parameter is assumed to be induced by a single-qubit
phase gate $U_{\phi }:=|g\rangle \langle g|+\exp (i\phi )|e\rangle \langle e|
$ acting on the qubit, where $\phi =\theta $ is some inference parameters
(see Fig.~1). To estimate the unknown parameter $\phi $ as precisely as
possible, the optimal input state may be chosen as $|\psi _{\mathrm{opt}%
}\rangle =(|g\rangle +|e\rangle )/\sqrt{2}$, which maximizes the QFI of the
output state $U_{\phi }|\psi _{\mathrm{opt}}\rangle $, see Ref.~\cite%
{Giovannetti2006}. In the following model, after the phase gate operation
and before the measurement performed, the qubit is assumed as an atom
coupled to a reservoir consisting of harmonic oscillators in the vacuum. The
total Hamiltonian of this typical model~\cite%
{BreuerBook,Garraway1997,Bellomo2007} reads
\begin{equation}
H=\omega _{0}\sigma _{+}\sigma _{-}+\sum_{k}\omega _{k}b_{k}^{\dagger
}b_{k}+(\sigma _{+}B+\sigma _{-}B^{\dagger })
\end{equation}%
with $B=\sum_{k}g_{k}b_{k}$, where $\omega _{0}$ denotes the transition
frequency of the atom with ground and excited states $|g\rangle $ and $%
|e\rangle $, and $\sigma _{\pm }$ the raising and lowering operators of
atom; $b_{k}^{\dagger }$ and $b_{k}$ are respectively the creating and
annihilation operators of the bath mode of frequencies $\omega _{k}$. $g_{k}$
denote the coupling constants. Then we consider Lorentzian spectral density $%
J(\omega )=\lambda W^{2}/\{\pi \lbrack (\omega _{0}-\omega )^{2}+\lambda
^{2}]\}$, where $W$ is the transition strength, and $\lambda $ defines the
spectral width of the coupling, which is related to the reservoir
correlation time scale $\tau _{\mathrm{B}}$ by $\tau _{B}=\lambda ^{-1}$~%
\cite{Garraway1997,BreuerBook}. The Lorentzian spectral density describe the
reservoir composed of lossy cavity, see Ref.~\cite{BreuerBook}. The time-local master equation of the
form (\ref{eq:K_t_Form}) can be obtained exactly as follows~\cite{BreuerBook}
\begin{equation}
\frac{\partial }{\partial t}\rho _{\mathrm{S}}(t)=\gamma (t)\left( \sigma
_{-}\rho _{\mathrm{S}}(t)\sigma _{+}-\frac{1}{2}\{\sigma _{+}\sigma
_{-},\rho _{\mathrm{S}}(t)\}\right)   \label{eq:ms_DJC}
\end{equation}%
where $\gamma (t)=-2\dot{h}(t)/h(t)$ with a crucial characteristic function ~%
\cite{BreuerBook}:
\begin{equation}
h(t)=%
\begin{cases}
e^{-\lambda t/2}\left[ \cosh \left( \frac{dt}{2}\right) +\frac{\lambda }{d}%
\sinh \left( \frac{dt}{2}\right) \right] , & W\leq \frac{\lambda }{2}, \\
e^{-\lambda t/2}\left[ \cos \left( \frac{dt}{2}\right) +\frac{\lambda }{d}%
\sin \left( \frac{dt}{2}\right) \right] , & W>\frac{\lambda }{2},%
\end{cases}
\label{eq:DJC_h}
\end{equation}%
where $d=\sqrt{|\lambda ^{2}-4W^{2}|}$.

Taking the initial state   $U_{\phi }|\psi _{\mathrm{opt}}\rangle $, the reduced
density matrix of the atom obeys  the master equation (\ref{eq:ms_DJC}).
Its solution is  $\rho _{\mathrm{S}}(t)=\left( I+\mathbf{B}\cdot \boldsymbol{\sigma} %
\right) /2$, where $\mathbf{B=(}h(t)\cos \phi ,-h(t)\sin \phi ,h(t)^{2}-1)$
and $\boldsymbol{\sigma} =(\sigma _{x},\sigma _{y},\sigma _{z})$. In order to calculate
the QFI flow, we first diagonalize this reduced density matrix as $\rho _{%
\mathrm{S}}(t)=\sum_{i}p _{i}(t)|\psi _{i}(t)\rangle \langle \psi _{i}(t)|$.
In this diagonal representation, the SLD with matrix elements $%
L_{ij}=2\langle \psi _{i}|\partial _{\phi }\rho_S |\psi _{j}\rangle /(p
_{i}+p _{j})$ is obtained explicitly as
\begin{equation}
L(t)=ih(t)\left[|\psi _{1}(t)\rangle \langle \psi _{2}(t)|-|\psi _{2}(t)\rangle \langle \psi
_{1}(t)|\right].
\end{equation}%
Further, we have $\mathcal{J}=-\mathrm{Tr(}\rho %
\left[ L,\sigma _{-}\right] ^{\dagger }\left[ L,\sigma _{-}\right]
)=-h(t)^{2}$. Then the exact solution for the QFI flow
\begin{equation}
\mathcal{I}_{\phi }(t)=\gamma (t)\mathcal{J}(t)=2h(t)\dot{h}(t).
\end{equation}%
is obtained, which leads to $\mathcal{F}_{\phi }=h(t)^{2}$.

\begin{figure}[!htb]
\begin{centering}
    \includegraphics[scale=0.25]{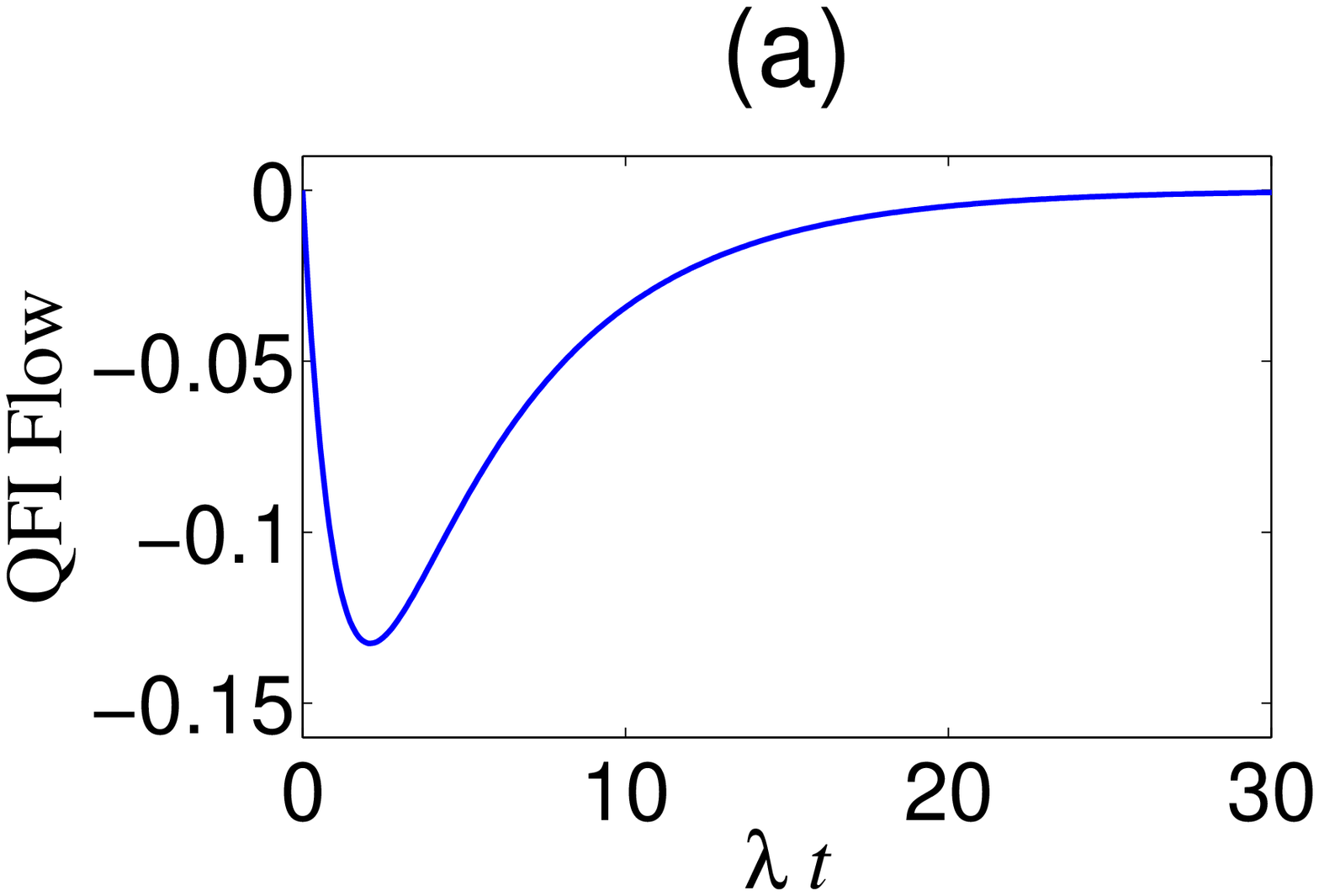}\includegraphics[scale=0.25]{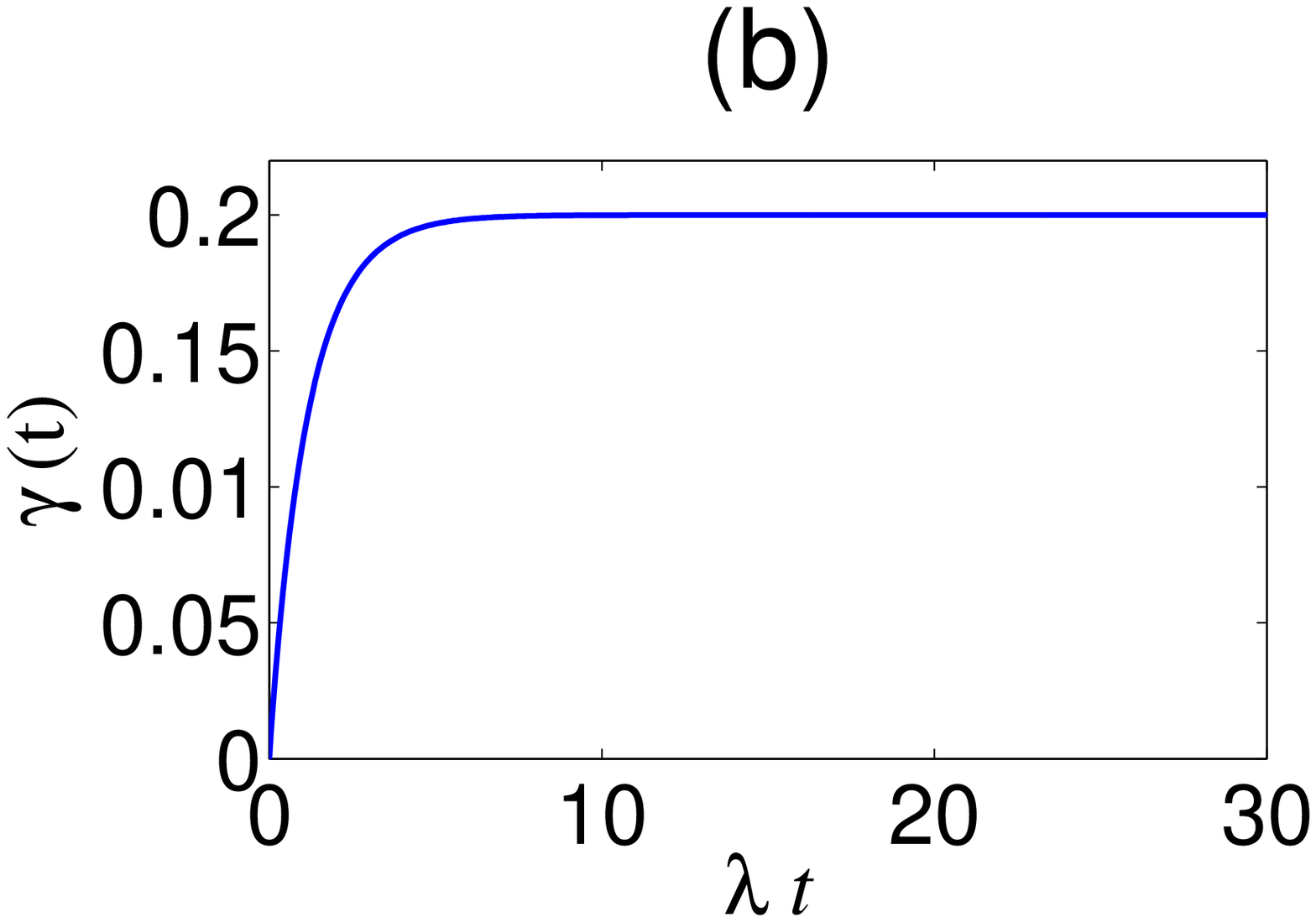}\\
    \includegraphics[scale=0.25]{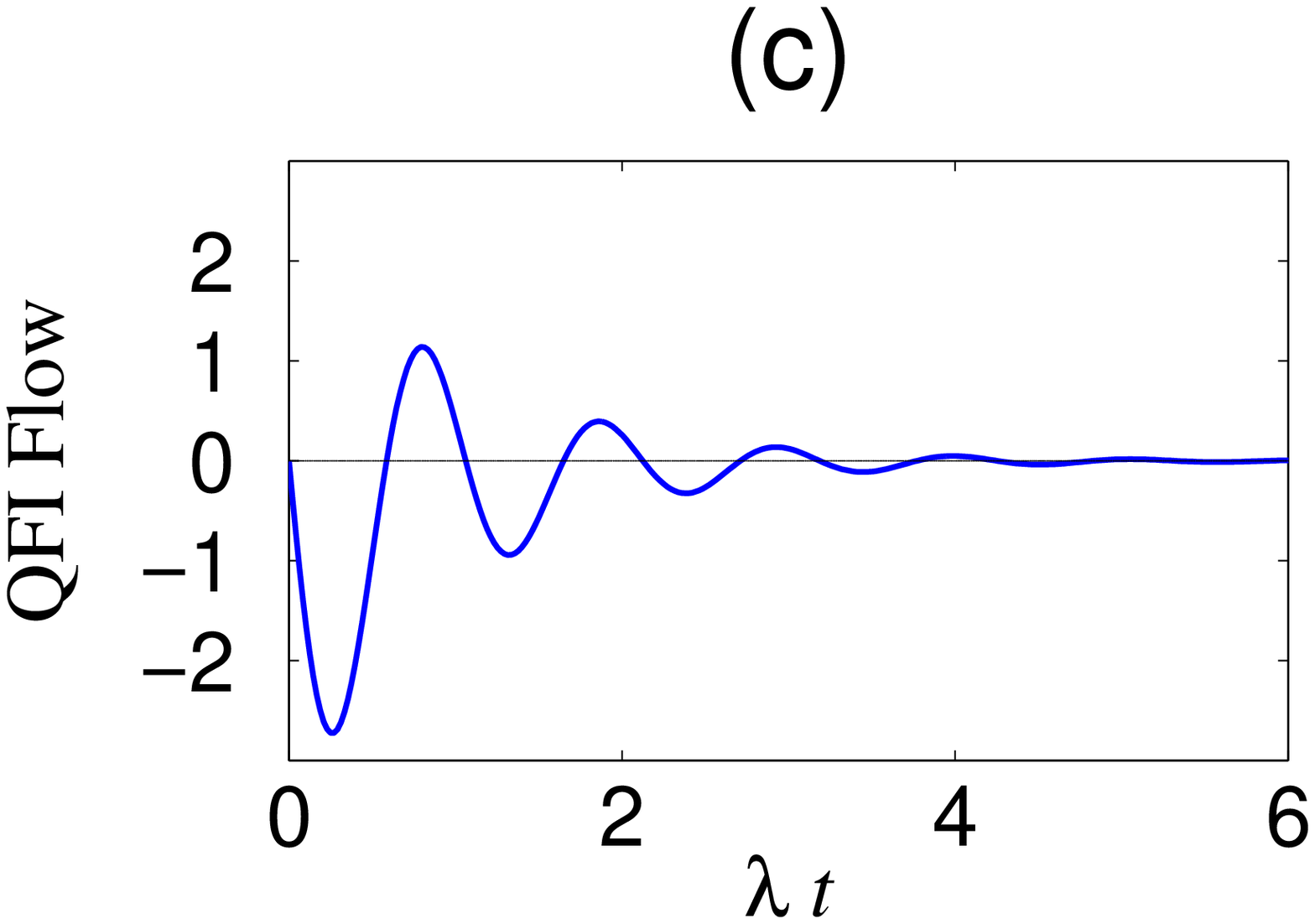}\includegraphics[scale=0.25]{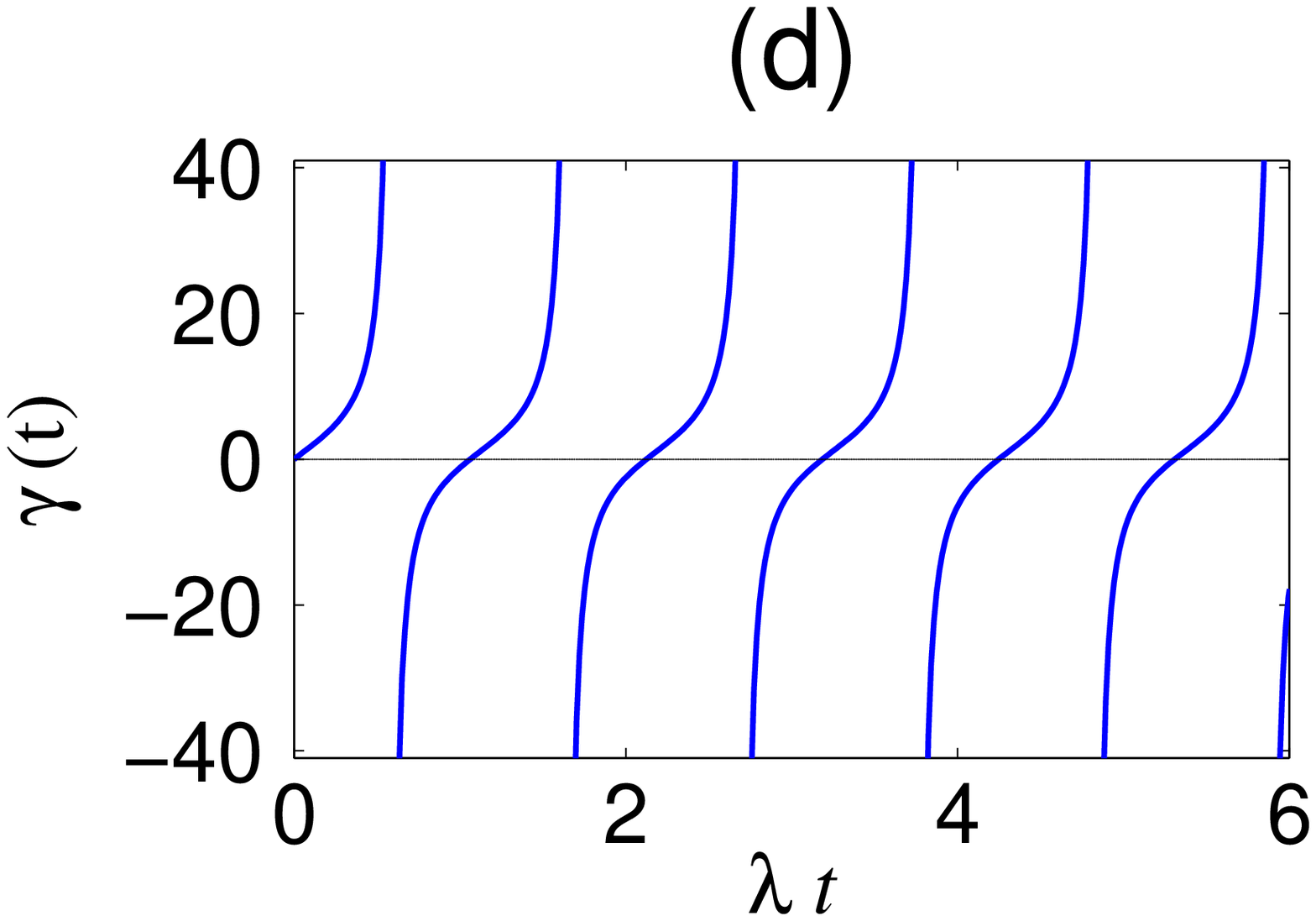}
  \end{centering}
\caption{Two-level atom coupled to reservoir with Lorentzian spectral
density: (a) QFI flow as a function of rescaled time, plotted in the weak
coupling regime ($W=0.3\protect\lambda$); (b) $\protect\gamma$ as a function
of rescaled time, $W=0.3\protect\lambda$; (c) QFI flow as a function of
rescaled time, plotted in the strong coupling regime ($W=3\protect\lambda$);
(d) $\protect\gamma$ as a function of rescaled time, $W=3\protect\lambda$. }
\label{fig:DJC}
\end{figure}

Therefore, the characteristic of the QFI flow is determined by the function $%
h(t)$, which has two very different kinds of behaviors. The corresponding properties
of the QFI flow are shown in Fig.~(\ref{fig:DJC}). In the weak coupling
regime ($W<\lambda/2$), the function $\gamma(t)$ is always positive, thus
the QFI is always lost during the time evolution of the open system. In the
strong coupling regime ($W>\lambda/2$), the function $\gamma(t)$ takes on
negative values within certain intervals of time, see Fig.~\ref{fig:DJC}
(d), which displays the non-Markovianity. Obviously in these time intervals,
the QFI flow is inward. It is remarkable that although $\gamma(t)$ diverges
at certain time, the QFI flow does not, see Figs.~~\ref{fig:DJC} (c) and
(d). This is because the QFI flow is determined by two factors, $\gamma(t)$
and $\mathcal{J}(t)$.

\section{Conclusion} In summary, based on the QFI flow, we
have proposed an information-theoretical approach for characterizing the
time-dependent memory effect of environment for its surrounding open quantum
systems. In this approach the Markovian process is considered as a QFI
erasure process, and the return of the QFI, i.e. an inward QFI flow, clearly
signatures the non-Markovian process. Using the time-local master equations,
we have showed that for each fixed time $t>0$ the QFI flow is decomposable
according to different dissipative channels, and the direction of each sub-flow
is determined by the sign of decay rates. With this decomposition form, the
relationship between the temporary appearance of negative decay rate and the
non-Markovian characteristic is justified. Although in the present work, the analysis
of the QFI flow is based on a time-local master equation, the concept of the QFI flow
may be still available in more general cases.

The present approach is associated with the current development of quantum
metrology, which concerns on finding an optimal fashion to make
high-resolution and highly sensitive measurement of physical parameters~\cite%
{Giovannetti2006}. Due to the interaction with environment in experiments,
like the photon losses in the optical interferometry or the presence of
quantum noise~\cite{noise}, the QFI will change and affect the precision of
the parameter estimation. Therefore, it is worthy to study the dynamical
evolution of the QFI in the context of quantum metrology, especially for non-Markovian
processes.

This work is supported by NSFC with Grant No.\,, 10874151, 10874091,
10935010; NFRPC with Grant No.\,2006CB921205; Program for New Century
Excellent Talents in University (NCET), and Science Fundation of Chinese
University.


\begin{thebibliography}{99}
\bibitem{BreuerBook} H.~P.~Breuer and F.~Petruccione, \emph{The Theory of
Open Quantum Systems} (Oxford University Press, Oxford, 2007).

\bibitem{ssb} J.~Q.~Liao, H. Dong, X. G. Wang, X.~F.~Liu, and C.~P.~Sun, e-print arXiv:0909.1230 (2009) and Refs. there
in.

\bibitem{Breuer2009} H.-P.~Breuer, E.-M Laine, and J. Piilo, \prl \textbf{103}, 210401 (2009).

\bibitem{Wolf2008} M. M. Wolf, J. Eisert, T. S. Cubitt, and J. I. Cirac, \prl  \textbf{101}, 150402 (2008).

\bibitem{Rivas2009} \'{A}.~Rivas, S. F. Huelga, and M. B. Plenio, e-print arXiv:0911.4270v1.

\bibitem{AlickiBook} R.~Alicki and K.~Lendi, \textit{Quantum Dynamical
Semigroups and Applications}, Lect. Notes Phys. 717 (Springer,  Berlin
Heidelberg, 2007).

\bibitem{Wootters1981} W.~K.~Wootters, Phys.~Rev.~D  \textbf{23}, 357
(1981).

\bibitem{Braunstein1994} S.~L.~Braunstein and C.~M.~Caves,  \prl
\textbf{72}, 3439 (1994).

\bibitem{Giovannetti2006} V.~Giovannetti, S. Lloyd, and L. Maccone, \prl
\textbf{96}, 010401 (2006).

\bibitem{Kossakowski2010} D.~Chru\'{s}ci\'{n}ski and A.~Kossakowski,  Phys.
Rev. Lett, \textbf{104}, 070406 (2010).

\bibitem{LindbladForm} V.~Gorini \textit{et al}.,  J. Math. Phys. \textbf{17}%
, 821 (1976); G.~Lindblad,  Commun. Math. Phys. \textbf{48}, 119 (1976).

\bibitem{Breuer2004} H.~P.~Breuer, Phys. Rev. A \textbf{70}, 012106  (2004).

\bibitem{TCLmethod} F. Shibata, Y. Takahashi, and N. Hashitsume, J. Stat. Phys. \textbf{17},
171 (1977); S.~Chaturvedi and  F.~Shibata, Z. Phys. B \textbf{35}, 297
(1979); F.~Shibata and  T.~Arimitsu, J. Phys. Soc. Jpn. \textbf{49}, 891
(1980); A.~Royer,  Phys. Rev. A \textbf{6}, 1741 (1972); A.~Royer, Phys.
Lett. A  \textbf{315}, 335 (2003).

\bibitem{FVmethod} R.~P.~Feynman and F.~L.~Vernon,  Ann. Phys. \textbf{24},
118 (1963); A.~O.~Caldeira and  A.~J.~Leggett, Physica A \textbf{121}, 587
(1983); B.~L.~Hu, J. P. Paz, and Y. Zhang, Phys. Rev. D \textbf{45}, 2843 (1992);
J.-H.~An, M. Feng, W.-M. Zhang, Quantum Inf. Comput. \textbf{9}, 0317 (2009).

\bibitem{Budini2006} A.~A.~Budini, \pra \textbf{74}, 053815 (2006).

\bibitem{estimationBook} C.~W.~Helstrom, \emph{Quantum Detection and
Estimation Theory} (Academic, New York, 1976), chap. VIII.4;  A.~S.~Holevo,
\emph{Probabilistic and Statistical Aspects of Quantum Theory}
(North-Holland, Amsterdam, 1982), especially Chaps. III.2  and VI.2.

\bibitem{negativeRate} J.~Piilo, S. Maniscalco, K. H\"{a}rk\"{o}nen, and K.-A. Suominen,  \prl \textbf{%
100}, 180402 (2008); H.-P. Breuer and  J. Piilo, Europhys. Lett. \textbf{85}
50004 (2009).

\bibitem{Wonderen2000} A.~J.~van Wonderen and K.~Lendi,  J. Stat. Phys.
\textbf{100}, 633 (2000).

\bibitem{Maniscalco2004} S.~Maniscalco, F. Intravaia, J. Piilo, and A. Messina, J. Opt. B:  Quantum
Semiclass. Opt. \textbf{6}, S98 (2004).

\bibitem{Wolf2008-1} M.~M.~Wolf and J.~I.~Cirac, Commun. Math. Phys.
\textbf{279}, 147 (2008).

\bibitem{Fujiwara2001} A.~Fujiwara, Phys.~Rev.~A, \textbf{63}, 042304
(2001).

\bibitem{Wootters1982} W.~K.~Wootters and W.~H.~Zurek, Nature  \textbf{299,}
802 (1982).

\bibitem{Garraway1997} B.~M.~Garraway, Phys. Rev. A \textbf{55}, 2290
(1997).

\bibitem{Bellomo2007} B.~Bellomo, R. Lo Franco, and G. Compagno, \prl
\textbf{99}, 160502 (2007).

\bibitem{noise} U.~Dorner, R. Demkowicz-Dobrzanski, B. J. Smith, J. S. Lundeen, W. Wasilewski, K. Banaszek, and I. A. Walmsley,  \prl \textbf{102},
040403 (2009); Y.~Watanabe, T. Sagawa, and M. Ueda, \prl \textbf{104},
020401 (2010).

\bibitem{ZYXu2010} Z.\ Y.\ Xu, W.\ L.\ Yang, and M.\ Feng, \pra \textbf{81}, 044105 (2010).

\end{thebibliography}
\end{document}